\documentclass[sigconf]{aamas} 

\usepackage{graphicx}
\usepackage{amsmath}
\usepackage{amsthm}
\usepackage[capitalize]{cleveref}
\usepackage{algorithm}
\usepackage{algorithmic}

\usepackage{wrapfig}
\usepackage{caption}
\usepackage{subcaption}

\DeclareMathOperator*{\esssup}{esssup}

\usepackage{xcolor}

\newcommand{\bk}[1]{} 
\usepackage[normalem]{ulem} 

\usepackage{balance} 

\setcopyright{ifaamas}
\acmConference[AAMAS '25]{Proc.\@ of the 24th International Conference
on Autonomous Agents and Multiagent Systems (AAMAS 2025)}{May 19 -- 23, 2025}
{Detroit, Michigan, USA}{A.~El~Fallah~Seghrouchni, Y.~Vorobeychik, S.~Das, A.~Nowe (eds.)}
\copyrightyear{2025}
\acmYear{2025}
\acmDOI{}
\acmPrice{}
\acmISBN{}

\acmSubmissionID{975}

\title{Bridging the Gap between Partially Observable Stochastic Games and Sparse POMDP Methods}

\author{Tyler Becker}
\affiliation{
  \institution{University of Colorado Boulder}
  \city{Boulder}
  \country{United States}}
\email{tyler.becker-1@colorado.edu}

\author{Zachary Sunberg}
\affiliation{
  \institution{University of Colorado Boulder}
  \city{Boulder}
  \country{United States}}
\email{zachary.sunberg@colorado.edu}

\begin{abstract}
Many real-world decision problems involve the interaction of multiple self-interested agents with limited sensing ability.
    The partially observable stochastic game (POSG) provides a mathematical framework for modeling these problems, however solving a POSG requires difficult reasoning over two critical factors: (1) information revealed by partial observations and (2) decisions other agents make.
    In the single agent case, partially observable Markov decision process (POMDP) planning can efficiently address partial observability with particle filtering.
    In the multi-agent case, extensive form game solution methods account for other agent's decisions, but preclude belief approximation. 
    We propose a unifying framework that combines POMDP-inspired state distribution approximation and game-theoretic equilibrium search on information sets. 
    This paper lays a theoretical foundation for the approach by bounding errors due to belief approximation, and empirically demonstrates effectiveness with a numerical example.
    The new approach enables planning in POSGs with very large state spaces, paving the way for reliable autonomous interaction in real-world physical environments and complementing multi-agent reinforcement learning.
\end{abstract}


\newcommand{\BibTeX}{\rm B\kern-.05em{\sc i\kern-.025em b}\kern-.08em\TeX}

\begin{document}


\pagestyle{fancy}
\fancyhead{}


\maketitle 


\section{Introduction}

This paper addresses the task of game-theoretic planning in large discrete, continuous, and hybrid state spaces with partial observability.
Interaction dynamics between agents is a significant source of uncertainty in planning.
This interaction can be cooperative~\citep[e.g.][]{lee2021disaster}, completely opposed (known as zero sum)~\citep[e.g.][]{becker2022imperfect}, or the agents may have more complex relationships with competing objectives that are not completely opposed or aligned (known as general sum)~\citep[e.g.][]{meta2022human}.
Aside from interaction uncertainty, we also consider state uncertainty since the complete state of an environment is rarely known and must instead be inferred with incomplete noisy and imperfect observations.

This combination of state and interaction uncertainty can result in complex behaviors that do not occur in single-agent settings.
For example, robots in cooperative data gathering or exploration settings may need to choose to actively share information to help the others complete the mission.
In the general sum autonomous driving setting, a self-driving car may need to reason about how other vehicles or pedestrians might react in order to successfully navigate difficult situations such as pulling into traffic.
Finally, in adversarial cases such as military settings, agents may choose to act less-predictably or bluff in order to confuse the other agents.

The partially observable Markov decision process (POMDP) is a flexible framework that handles state uncertainty but not interaction uncertainty. Many planning methods, both online~\citep[e.g.][]{silver2010pomcp,sunberg2018pomcpow} and offline~\citep[e.g.][]{kurniawati2008sarsop}, have been developed for solving POMDPs efficiently in practice.
On the other hand, approaches from game theory focus mainly on interaction uncertainty and handle state uncertainty less robustly because they  plan on the history space~\citep[e.g.][]{zinkevich2007cfr,moravcik2017deepstack}. 
Some deep reinforcement learning (RL) methods~\citep[e.g.][]{vinyals2019grandmaster,jaderberg2019human} have enjoyed great success in environments with interaction and state uncertainty.
Our tree search planning approach is complementary to deep RL in several ways.
For example, tree search can be used as an improvement operator during the training process~\citep{silver2018alphazero,moss2024betazero}; a tree search algorithm can act as a backbone to combine learned models for a compositional approach to planning~\citep{deglurkar2023compositional}; or the structure proposed in this work could be used as an algorithmic prior for reinforcement learning algorithms~\citep{jonschkowski2018differentiable}.

The specific contributions of this paper are the following: First, after a review of background material in \cref{sec:background}, \cref{sec:cdit} defines the conditional distribution information set tree (CDIT) structure which can accommodate both particle-based belief approximation and multi-agent reasoning.
Next, \cref{sec:equilibria} describes a counterfactual regret minimization (CFR) algorithm that finds a Nash equilibrium in the zero-sum case by exploring important parts of the CDIT.
\Cref{sec:convergence} then provides theoretical justification for the CDIT approach showing that if a Nash equilibrium is calculated using the CDIT approximation, it is likely close to a Nash equilibrium for the true game. Notably, the theoretical properties have no direct computational complexity dependence on the size of the state space.
Finally, \cref{sec:experiments} contains a numerical experiment showing that the algorithm described in \cref{sec:equilibria} can efficiently find effective strategies in a tag game on a continuous state space, a task that is impossible for existing methods based on POMDPs or extensive form games.

\section{Background}\label{sec:background}


\subsection{Partially observable stochastic games}

The partially observable stochastic game (POSG), also called the partially observable Markov game (POMG), is a mathematical formalism for problems where multiple agents make decisions sequentially to maximize an objective function~\citep{albrecht2023marl, kochenderfer2019algorithms}.
A particular finite horizon POSG instance is defined by the tuple $(\mathcal{N}, \mathcal{S}, \mathcal{A}, \mathcal{O}, \mathcal{T}, \mathcal{Z}, r, D, \gamma, b_0)$, where
$\mathcal{N}$ is the set of players ($i \in \mathcal{N}$) playing the game.
$\mathcal{S}$ is a set of possible states;
\bk{ $\mathcal{A}$ is the cartesian product of $\mathcal{A}^i$ where } $\mathcal{A}^i$ is a set of player $i$'s actions;
$\mathcal{O}^i$ is a set of player $i$'s observations;
$\mathcal{T}$ is the state transition function where $\mathcal{T}(s' | s, a)$ is the probability of transitioning from \bk{joint?} state $s$ to \bk{joint?} state $s'$ via joint action $a$; 
$\mathcal{Z}^i$ is the observation function for player $i$ where $\mathcal{Z}^i(o^i|s,a, s')$ is the probability that player $i$ receives observation $o^i$ given a transition from state $s$ to state $s'$ via joint action $a$;
$r^i(s, a)$ \bk{$R^i$?} is the scalar reward function for player $i$, given state $s$ and joint action $a$;
$D \in \mathbb{N}$ is the number of time steps in the horizon; \bk{D is the horizon of the problem or the set of timesteps?}
$\gamma \in [0,1]$ is the discount factor; and
$b_0$ is the initial distribution over states.
The results in this paper apply to finite horizon problems. However, they can be extended to infinite-horizon discounted problems with bounded reward by choosing a planning horizon large enough that subsequent discounted rewards will be sufficiently small.
\bk{What is the relationship between $\mathcal{O}^i$ and $\mathcal{O}$, etc. ...}

\bk{$b_0$ in $h$?}

A policy for player $i$, $\pi^i$, is a mapping from that player's action-observation history, $h^i$, to a distribution over actions. Similarly, we define $\sigma \in \Sigma$ to a pure strategy that maps from a player's action-observation history to a single action.
The space of possible polices for agent $i$ is denoted $\Pi^i$.
A joint policy, $\pi = (\pi^1, \pi^2, ...)$, is a collection of individual policies for each player.
The superscript $-i$ is used to mean ``all other players''. For example $\pi^{-i}$ denotes the policies for all players except $i$.
Player $i$'s objective is to choose a policy to maximize his or her utility,
\begin{equation}
    U^i(\pi) = \mathbb{E}_{\mathcal{T},\mathcal{Z},\pi}
    \left[
        \sum_{t=0}^D \gamma^t r^i(s_t, a_t)
    \mid s_0 \sim b_0\right]\,.
\end{equation}
Since this objective depends on the joint policy, and the reward functions for individual players may not align with each other, POSGs are not optimization problems where locally or globally optimal solutions are always well-defined.
Instead of optima, there are a variety of possible solution concepts.
The most common solution concept and the one adopted in this paper is the Nash equilibrium.
A joint policy is a Nash equilibrium if every player is playing a best response to all others.
Mathematically a best response to $\pi^{-i}$ is a $\pi^i$ that satisfies
\begin{equation} \label{eq:br}
    U^i(\pi^i, \pi^{-i}) \ge U^i(\pi^{i\prime}, \pi^{-i})
\end{equation}
for all possible policies $\pi^{i\prime}$. In a Nash equilibrium, \cref{eq:br} is satisfied for all players.

One particularly important taxonomic feature for POSGs is the relationship between the agents' reward functions.
In \emph{cooperative} games, all agents have the same reward function,
$r^i(s, a) = r^j(s, a)$ for all $i$ and $j$.
In a two player \emph{zero-sum} game, the reward functions of the two players are directly opposed and add to zero, $r^1(s, a) = -r^2(s, a)$.
When there are no restrictions on the reward function, the term \emph{general-sum} is used to contrast with cooperative, zero-sum, or other special classes.
The CDIT structure in \cref{sec:cdit} and the analysis in \cref{sec:convergence} are applicable to the general-sum case, while the algorithm in \cref{sec:equilibria} applies only to the zero-sum case.

\subsection{Particle filtering and tree search for POMDPs}

A POMDP is a POSG special case where there is only one player. Hence, a POMDP can be described by the same tuple as a POSG, with only one action set, observation set, and conditional observation distribution for the sole player.
Tree search is a scalable approach for solving POMDPs~\cite{silver2010pomcp,ye2017despot,sunberg2018pomcpow,garg2019despotalpha}.

For POMDPs, each node in the planning tree corresponds to a history, and actions are typically chosen by estimating the history-action value function, $Q(h,a)$.

A POMDP is equivalent to a belief MDP, where a belief is the state distribution conditioned on the action-observation history ($b(s) = P(s \mid h)$).
As the belief is a sufficient statistic for value~\cite{hauskrecht2000value}, the value function can be conditioned on beliefs instead of full histories ($Q(b,a)$). 
While exact Bayesian belief updates work for well for solving relatively small POMDPs~\cite{kurniawati2008sarsop}, these methods suffer from the curse of dimensionality in large state and observation spaces. 
If a weighted particle filter~\cite{kochenderfer2019algorithms} and sparse sampling~\cite{kearns2002sparse} are used to approximate the belief MDP, there need be no direct computational dependence on the size of the state or observation spaces, breaking the curse of dimensionality in the state and observation spaces~\cite{lim2023optimality}.

\subsection{Imperfect information extensive form games} \label{sec:EFGs}
Imperfect information extensive form games (EFGs) are very similar to POSGs as they both provide a model for multiagent interaction under imperfect state information. However, there are three structural differences: First, EFGs assume a sequential nature to the interaction rather than the simultaneous interaction of POSGs. Second, rather than defining a state-based reward, EFGs define rewards only for terminal histories. Finally, instead of reasoning about observations, EFGs may arbitrarily group histories into information sets wherein a player is unable to distinguish between these histories.
While it is theoretically possible to represent every POSG as an EFG by representing all sources of uncertainty as chance plays, this is unnatural for games designed to represent real world physical scenarios.

A method for finding Nash equilibria in zero-sum EFGs is through regret matching over the pure policy space~\cite{hart2000regret-matching}. The regret in some pure strategy $\sigma$ is defined as the utility that could have been gained by playing strategy $\sigma$ instead of the policy that was actually played $\pi_i^t$. Mathematically, this is given by
\begin{equation} \label{eq:regret}
    R_i^T(a) = \sum_{t=1}^T [U(a, \pi_{-i}^t) - U(\pi_{i}^t, \pi_{-i}^t)]\,. 
\end{equation}

The regret matching strategy is then defined as one that plays pure strategies with probability proportional to the positive regret ($R_i^{T, +}(\sigma) = \max\{R_i^T(\sigma),0\}$) in not playing them in the past i.e. 
\begin{equation}
    \pi_i^{T+1}(\sigma) = \begin{cases}
        \frac{R_i^{T, +}(\sigma)}{\sum_{a \in \Sigma^i} R_i^{T, +}(\sigma)} \quad & \text{if } \sum_{a \in \Sigma^i} R_i^{T, +}(a) > 0 \\
        1 / |\Sigma^i| & \text{otherwise} \, .
    \end{cases}
\end{equation}
It is shown that the average of these regret matching strategies converges to a Nash equilibrium in zero-sum games~\cite{hart2000regret-matching}.

However, with EFGs the full policy space is often too large too enumerate completely. So, for these EFGs, counterfactual regret minimization~\citep{zinkevich2007cfr} can be employed that performs the same regret matching and policy averaging procedure but over counterfactual regrets in each action for each decision point. The counterfactual regret for a decision point is simply the regret (\cref{eq:regret}) weighted by the probability of reaching that information set should the player updating their strategy play strictly to reach this decision point. This method of solving EFGs has been used as a base algorithm for solving extremely large games such as heads up no limit Texas hold 'em poker~\citep{moravcik2017deepstack,brown2018superhuman}.


\subsection{Limitations of POMDP approaches}

In physical domains such as robotics or aerospace, both POMDPs and EFGs have severe limitations.
POMDP approaches are popular in robotics \citep{lauri2022partially,kurniawati2022partially}, and have enjoyed success due to the many efficient methods for approximating the belief such as Kalman filtering or particle filtering.
However, POMDPs are fundamentally single agent optimization problems rather than games.
Cooperative POSGs, sometimes called \emph{decentralized POMDPs} \citep{oliehoek2016concise}, can be solved using POMDP-based techniques \citep[\textit{e.g.}][]{zhang2019online} since all agents seek the same goal.
However, finding certain types of equilibria in general-sum games is fundamentally impossible.
A simple example is a mixed Nash equilibrium:
since POMDPs always have at least one deterministic optimal policy, it is impossible for a POMDP-based approach to find an equilibrium in a game that only has a mixed Nash equilibrium.
This also means that recursive POMDP-based interaction approximations such as I-POMDPs \citep{doshi2009monte} cannot calculate these equilibria.

Moreover, while beliefs, defined in this context as the conditional distribution of the state given the action-observation history, are fundamentally important in POMDPs because they form the basis for many dynamic programming algorithms, they can not be used in the same way for POSGs.
In a POSG, the state distribution conditioned on an agent's private history also depends on the other agents' policies, so it is impossible to calculate belief distributions without knowing these policies first.
Since the very goal in a POSG is to find a joint policy that is an equilibrium, the other agents' policies are not known a-priori, and it is not straightforward to use POMDP beliefs in intermediate steps to find POSG equilibria.
Existing methods that have attempted to resolve this require exact belief updates and are therefore limited to beliefs that can be represented with a small number of parameters such as categorical distributions over small discrete state spaces~\cite{delage2023hsvi}. 

\begin{figure*}[htb!]
    \centering
    \includegraphics[width=0.75\textwidth]{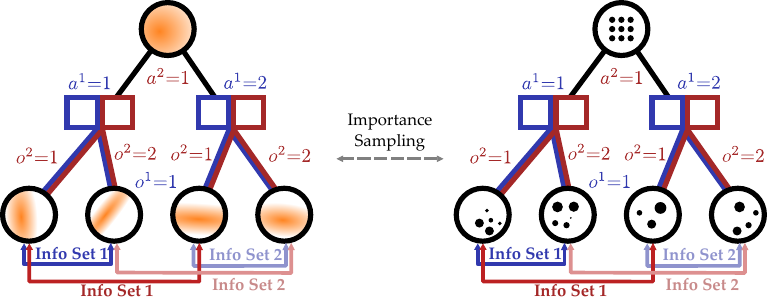}
    \caption{Illustration of a CDIT (left) and its particle approximation (right) for a POSG with $\mathcal{A}^1$=$\mathcal{O}^2$=\{1,2\}, $\mathcal{A}^2$=$\mathcal{O}^1$=\{1\}.}
    \label{fig:cdit}
\end{figure*}

\subsection{Limitations of EFG approaches}

Instead of using belief distributions, EFG-based approaches represent state uncertainty by grouping states 
into information sets without assigning probability to the set members.
By deferring this probability assignment, algorithms such as CFR can find equilibria by adjusting all agents' policihes simultaneously.
While the EFG formalism is general -- POSGs can be represented as EFGs by modeling all uncertainty as chance player moves -- it was designed for relatively simple tabletop games, and it is extremely inconvenient to model physical world domains as EFGs. 
More importantly, EFGs cannot easily incorporate belief approximation techniques such as Kalman filtering or particle filtering, which have been crucial in handling uncertainty in physical domains over the last half century.

Focus has since shifted from EFGs as a formalism to factored observability stochastic games (FOSGs) which are equivalent to POSGs save the factorization of observations into public and private observations. Private observations are known only to an individual player that receives their private observation, and public observations are received by all players. This additional assumption allows for the notion of public states that narrow the number of histories required for subgame solving. However, while much work has been done to handle larger history spaces, little work has accounted for large or even continuous world state spaces. It is commonly assumed that rewards and transition probabilities can be ascertained directly from the history. Even if the underlying state space is considered, these required quantities are assumed to exactly attainable via discrete Bayesian filtering. 
When considering this underlying state space, the history-reward is given by 
\begin{equation} \label{eq:exact-hist-reward}
    r(h,a) = \int_\mathcal{S} P(s | h) r(s,a) ds\,,
\end{equation}
and the history transition density is given by 
\begin{equation} \label{eq:exact-hist-transition}
    p(h'=hao|h,a) = \int_{\mathcal{S}} Z(o \mid a, s') \left[\int_{\mathcal{S}} T(s' \mid s, a)\, p(s|h) ds \right] ds'\, .
\end{equation}
Without further restrictive assumptions on the structure of the problem, these quantities are generally intractable to calculate.

The work on scaling existing game theoretic solvers fall into several categories: one category ~\cite{hu2021learned,fickinger2021scalable, sokota2022fine, vsustr2021particle, tian2020joint, moravvcik2017deepstack, zarick2020unlocking, brown2020combining, schmid2023student} uses function approximation via neural network as a key component. Our work is a complementary alternative because we do not rely on deep learning though it has potential to be combined with such techniques in the future. Training neural network models can be difficult and may require large amounts of data, and in theoretical analysis, it is often assumed that neural networks will adequately approximate some function when convergence of the neural network is not guaranteed. These deep learning methods generally use tabular methods with convergence guarantees as a foundation and then leverage the generalization ability of deep learning to scale to larger spaces. While empirically performant, the incorporation of deep function approximation generally breaks the guarantees established by the underlying tabular algorithm.


Another category \cite{solinas2024history, liu2023opponent, Libratus, brown2017safe, delage2024hsvi, sokota2023abstracting} requires explicit reach probability distributions and/or does not account for a state space underlying the history space. As demonstrated in \cref{eq:exact-hist-reward,eq:exact-hist-transition}, these methods break when extending to very large or continuous state spaces.


\subsection{Summary of Improvements}

Our approach provides a straightforward and theoretically grounded path to convergence in large or continuous state spaces without the need for extensive hyperparameter optimization. The proposed methodology efficiently approximates expected utilities which would be impossible to calculate exactly in the general case. Moreover, by using methods with established probabilistic convergence proofs that are not directly dependent on the size of the state space, we offer theoretical guarantees for the performance of our algorithm.




\section{Conditional distribution information set trees} \label{sec:cdit}


In order to overcome the limitations above, we define a new tree structure called the conditional distribution information set tree (CDIT) that combines history-conditioned state distributions, similar to POMDP beliefs, with information sets similar to those in EFGs. A CDIT is shown in \cref{fig:cdit}, which will be explained in the following sections.

\subsection{Joint conditional distribution trees}

The base structure of a CDIT is a joint conditional distribution tree consisting of alternating layers of joint action nodes (rectangles in \cref{fig:cdit}) and joint observation nodes (unfilled circles in \cref{fig:cdit}).
The history for a node is the sequence of joint actions and joint observations on the path to that node from the root.
Each depth $d$ observation node has an associated state distribution, $b_d$, conditioned on the history up to that point, that is
\begin{equation}
    b_d(s_d) = P(s_d \mid h_d) = P(s_d \mid b_0, a_0, o_1, ... a_{d-1}, o_d) \text{.} 
\end{equation}
This distribution can be calculated exactly using Bayes' rule: 
\begin{equation}
    b_d(s_d) \propto Z(o_d \mid a_{d-1}, s_d) \int_{\mathcal{S}} T(s_d \mid s_{d-1}, a_{d-1})\, b_{d-1}(s_{d-1})ds_{d-1}\, ,
\end{equation}
where $a_{d-1}$ is the joint action for the parent node and $b_{d-1}$ is the joint conditional distribution for the grandparent node.
However, CDITs are most scalable when this belief is approximated. 
Any approximation, for example an extended Kalman filter or Gaussian mixture model can be used, but this work focuses on a \emph{particle CDITs}, where each distribution is represented by $C$ particles (filled circles in \cref{fig:cdit}):
\begin{equation} 
    \bar{b}_d = \{(s_{d,i}, w_{d,i})\}_{i=1}^C .
\end{equation}
Particles are propagated by sampling the joint transition distribution
    $s_{d+1, i} \sim \mathcal{T}(s_{d+1, i} \mid s_{d,i}, a)$,
and particle weights are updated according to observation probability:
    $w_{d+1, i} = w_{d, i} \cdot \mathcal{Z}(o \mid a, s_{d+1,i})$.

%

\subsection{Combining joint conditional distributions with information sets}

Since the distributions in the tree described above are conditioned on joint histories, they contain more information than any one player has at a given depth.
In order to limit the information that policies can be conditioned on, distribution nodes corresponding to histories that are indistinguishable to a player are grouped together into information sets for each player. 
Specifically, two joint observation nodes are in the same information set for player $i$ if all of the actions and observations for player $i$ in the history leading up to that node are identical.
This grouping is similar to the information set concept in EFGs.
However, while EFGs may arbitrarily group states into information sets, CDITs by definition group nodes according to the criterion above.

The combination of a joint conditional distribution tree and history-based information sets constitutes a CDIT.
\Cref{fig:cdit} shows the first two layers of a particle CDIT for a simple example.
The root node contains particles sampled from $b_0$.
Since Player 1 has two actions and Player 2 only has one action, there are two joint action nodes, and since Player 1 has only one observation and Player 2 has two observations, there are two observation children for each action node.
This yields four state distributions in the bottom layer.
At this level, each player has two information sets.
For Player 1, the left two nodes are grouped into an information set corresponding to $(a^1 = 1, o^1 = 1)$, and the right two are grouped into a set corresponding to $(a^1 = 2, o^1 = 1)$.
For Player 2, the first and third nodes are grouped into the $(a^2=1, o^2=1)$ set and the second and fourth are grouped into the $(a^2=1, o^2=2)$ set.

An information set in a CDIT can be interpreted as a summary of the information about the state implied by the history observed by an agent without assuming anything about other agents' policies.
It is not a distribution, however it does contain some information about the likelihood of states.
In particular, if the policies of the other agents, $\pi^{-i}$, are known, then the state distribution can be calculated by weighting each constituent conditional distribution in the information set by the likelihood of its associated joint history under $\pi^{-i}$. 
This also implies that the probability of a given state is at least as high as the minimum probability of that state in all of the constituent conditional distributions and no more than the maximum probability of that state in all of those distribution.
Thus, even without knowing $\pi^{-i}$, if a state has high probability in all of the distributions in an information set, the agent can conclude that that state is likely, and conversely, if it has low probability in all of the distributions, it is unlikely.

\section{Finding approximate Nash equilibria on CDITs for zero sum games} \label{sec:equilibria}


While the CDIT provides a suitable structure for approximating a POSG, an equilibrium-finding algorithm is needed to solve it.
This section describes a particular algorithm, external sampling counterfactual regret minimization (ESCFR) that can find Nash equilibria for zero sum games. 

ESCFR traverses a constructed particle CDIT by recursively searching all actions of the traversing player, but only sampling from the opposing player's strategy as well as the transition densities of the environment. The allure of external sampling comes from the theoretical asymptotic convergence analysis wherein ESCFR converges in regret with $O(1/\sqrt{T})$ for $T$ iterations which is equivalent to the asymptotic convergence rate of vanilla CFR. However, the asymptotic computational cost of each ESCFR iteration is $O(\sqrt{|H|})$, whereas the asymptotic computational cost of vanilla CFR is $O(|H|)$, where $|H|$ is the size of the searched joint history space. This makes ESCFR more computationally efficient than vanilla CFR.
Furthermore, while we could sample from everything, resorting to outcome sampling CFR (OSCFR), this method introduces an importance sampling term in the computed regret that yields high variance samples and induces numerical instability in larger games.

Counterfactual regret minimization (CFR) algorithms reason using counterfactual action utilities $q_{i,c}$ for information sets $I$:
\begin{equation}
    q_{i,c}^\pi(I,a) = \sum_{h \in I}P^\pi_{-i}(h)q_i^\pi(h,a),
\end{equation}
where $P^\pi_{-i}(h)$ is the counterfactual reach probability of history $h$ for player $i$, and $q_i^\pi(h,a)$ is the non-counterfactual history-action value. Counterfactual reach probability stems from the ability to factor reach probability into player components i.e. $P^\pi(h) = P^\pi_{i}(h) \cdot P^\pi_{-i}(h)$. The counterfactual reach probability can then be interpreted as the probability that history $h$ is reached supposing that all players play according to the strategy $\pi$ except player $i$ who plays to reach $h$.
The history-action value is defined recursively as follows:
\begin{equation}
    q_i^\pi(h,a) = \mathbb{E}_{\pi^{-i},\mathcal{T},\mathcal{Z}}\left[r^i(h,a) + \gamma v^\pi_i(h'=hao)\right] \text{,}
\end{equation}
\bk{Should it be $\pi^{-i}$ under expectation?}
where $v_{i}^\pi(h) = \mathbb{E}_{a \sim \pi}\left[ q_{i}^\pi(h,a)\right]$.

However, for particle filter CDITs and general large-state space POSGs alike, the counterfactual reach probabilities $P^\pi_{-i}(h)$ become very difficult to compute. These quantities depend on players' policies $\pi_i$ as well as environmental factors $(\mathcal{T}, \mathcal{Z})$. As such, the probability of transitioning from history $h$ to history $h' = hao$ is given by
\begin{equation}
    P^\pi(h'=hao | h) = p(o | h, a) \cdot \Pi_{i \in \mathcal{N}}\pi_i(I_i(h), a_i)
\end{equation}
\bk{Consider $\pi_i(a_i|I_i(h))$?}
Here $p(o \mid h, a)$ is analogous to the ``chance player'' policy for EFGs. In a particle CDIT, the probability of an observation given some particle belief $\bar{b}_d$ and joint action $a$ is
\begin{equation} \label{eq:belief-mdp-transition-integral}
    p(o \mid \bar{b}_d,a) = \frac{\sum_{i=1}^C w_{d,i} \cdot [\int_S \mathcal{Z}(o \mid a, s')\mathcal{T}(s' \mid s_{d,i}, a)ds']}
    {\sum_{i=1}^C w_{d,i}} .
\end{equation}

However, computing this quantity is generally very difficult or even impossible.
This necessitates external sampling~\cite{lanctot2009mccfr} where a sampled regret $\tilde{r}$ is accumulated
\begin{equation}
    \tilde{r}^i(I,a) = (1 - \pi(I,a)) \sum_{h \in Q_I} q^\pi_i(h,a),
\end{equation}
where $Q_I$ denotes the subset of histories in information state $I$ that have been sampled.
As such, no reach probabilities need to be explicitly recorded, sidestepping the necessity to compute intractible integral given in \cref{eq:belief-mdp-transition-integral}. Similar to CFR outlined in \cref{sec:EFGs}, sampled regrets are accumulated over iterations and policy iterates play actions with probability proportional to the regret in not having played them in the past. 



\section{Convergence guarantees for approximate Nash equilibria on CDITs} \label{sec:convergence}

When using a particle CDIT to approximate a game, a crucial question is whether equilibria computed on the CDIT, for example with the ESCFR algorithm above, converge to equilibria in the original game as the number of particles increases.
This section answers that question affirmatively.
Although the bounds in this section are extremely loose, they show that the approach is sound.
Moreover, unlike the previous section which focused on zero-sum games, these results apply to general sum games, and the bounds have no direct dependence on the size of the state space, suggesting that the general approach can scale to large state spaces.

We separate the convergence guarantees into three parts. First, we show that the suboptimality of a solution calculated using an approximate game is bounded when applied to the true game in \cref{sec:approx-game-delta}. Then, we bound utility approximation error of this approximate game in \cref{sec:proof-policy-eval-error}. 
Next, we show that using a sampled subset of the strategies and observations in the approximate game is sufficient to solve the approximate game in \cref{sec:proof-no-tree-equiv-incentive}.
Finally, we bound the suboptimality of a sparse ESCFR solution in \cref{sec:proof-final-convergence}.

\subsection{Suboptimality of Approximate Games} \label{sec:approx-game-delta}

A CDIT yields an approximate estimate of the utility function $\hat{U}(\pi)$.
Since this estimate differs from the true utility, $U(\pi)$, we must characterize how equilibria computed with $\hat{U}(\pi)$ are related to equilibria on the original game.
To do this, we use a normal form game where each action corresponds to a pure policy for a subset of the players in the original POSG.
Specifically, this game is defined by a set of $|\mathcal{N}|$ payoff matrices, $A$, with each player's payoff matrix denoted $A^i$.
The payoff matrix entries for the true and approximated games are defined as
\begin{equation}
A^i_{j,k} \equiv U^i(\sigma^i_j, \sigma^{-i}_k) \qquad \text{and} \qquad \hat{A}^i_{j,k} \equiv \hat{U}^i(\sigma^i_j, \sigma^{-i}_k)
\end{equation}
where $\sigma^i_j$ is the $j$th pure policy played by player $i$ and $\sigma^{-i}_k$ is the $k$th joint policy between all players that are not player $i$. 
By considering mixtures of the pure strategies in this normal form game, we consider all mixed strategies in the POSG since POSGs inherently have perfect recall and thus Kuhn's theorem applies~\cite{kuhn1950extensive}. 

A Nash equilibrium requires that no player has incentive to deviate from their current strategy. For a game with payoff matrices $A$, we denote the value of the incentive to deviate with
\begin{equation}
\begin{aligned}
    \delta^i_A \equiv \max_{\pi^{i\prime}}\left(\pi^{i\prime\,T} A^i \pi^{-i}\right) - \pi^{i\,T} A^i \pi^{-i} .
\end{aligned}
\end{equation}
The sum of deviation incentives for all players,
\begin{equation}
    \textsc{NashConv}_A(\pi) = \sum_i \delta^i_A(\pi) \text{,}
\end{equation}
will be used as a distance metric between the current strategy and a Nash equilibrium \citep{johanson2011accelerating}.
The matrix of approximation errors will be denoted with $E^i \equiv A^i - \hat{A}^i$.
Using this notation, the following Lemma bounds the error in \textsc{NashConv} given payoff estimation errors.

\begin{lemma} \label{lemma:nashcov-bound}
    In a game with payoff matrices $A$,
    the deviation incentive for Player $i$ from a policy $\pi$ can be upper bounded by
    \begin{equation}
    \begin{aligned}
        \delta^i_{A}(\pi) &\le \delta^i_{\hat{A}}(\pi) + \delta^i_{E}(\pi) \\
        &\le \delta^i_{\hat{A}}(\pi) + 2||E^i||_\infty\, \text{,}
    \end{aligned}
    \end{equation}
    and
    \begin{equation} 
    \begin{aligned}
        \textsc{NashConv}_A(\pi) & \le \sum_i \left[\delta^i_{\hat{A}^i}(\pi) + 2||E^i||_\infty\right] .
    \end{aligned}
    \end{equation}
\end{lemma}
The proof for \cref{lemma:nashcov-bound} is in the appendix. The proof relies on the ability to upper bound best response utility in normal form by the sum of maximizations upper bounding a maximization over a sum.

\subsection{Particle CDIT Policy Evaluation Error} \label{sec:proof-policy-eval-error}

%
%
%

We now turn our attention to bounding the error in the particle CDIT utility estimate, $\hat{U}$.
It should be noted that evaluating a joint policy on the joint conditional distribution tree core of a CDIT is functionally identical to evaluating a POMDP policy.
As such, we can invoke recently-developed utility approximation error bounds for particle filter approximation in POMDPs. Specifically, we use the sparse sampling-$\omega$-$\pi$ tree\citep{lev2024simplifying} to sparsely sample the state and history space to estimate the value of an arbitrary deterministic policy:

\begin{theorem} \label{theorem:single-policy-error-bound}
For all policies $\sigma, t = 0, . . . , D$, the following bounds hold with probability of at least $1 - 5(4C)^{D+1}\exp(-C \cdot \acute{k}^2)$:
\begin{equation}
    \left|U^{i}(\sigma)-\hat{U}^{i}(\sigma)\right| \leq \epsilon_{\omega\pi},
\end{equation}
where
\begin{equation}
    \epsilon_{\omega\pi} = \lambda\left[
    2\left(\frac{1-\gamma^{D+1}}{1-\gamma}\right) - 1
    \right]
\end{equation}
\begin{equation}
    k_{\max }(\lambda, C)=\frac{\lambda}{4 V_{\max,i} d_{\infty}^{\max }}-\frac{1}{\sqrt{C}}>0,
\end{equation}
\begin{equation}
    \acute{k}=\min \left\{k_{\max }, \lambda / 4 \sqrt{2} V_{\max }\right\} .
\end{equation}
\end{theorem}

\begin{proof} (Sketch)
    This follows immediately from Theorem 3 of \citet{lev2024simplifying} applied to the root node of a joint conditional distribution tree at $t=0$ using a deterministic joint policy $\sigma$. A more detailed account is given in the appendix.
\end{proof}

\subsection{Surrogate Game Sufficiency} \label{sec:proof-no-tree-equiv-incentive}
For a particle CDIT, because we are sparsely sampling the history space, we need only consider a subset of the full policy space.  We call this restricted subspace of the full policy space the "surrogate" policy space, denoted by $\Sigma_s$.
We consider two policies "tree-equivalent" if they take all of the same actions in all beliefs sampled in the tree. 
More formally, sparse sampling-$\omega$-$\pi$  constructs a tree that samples a subset of the total history space $\mathcal{H}_{\omega\pi} \subseteq \mathcal{H}$. Any two deterministic policies $\sigma^A, \sigma^B$ are tree-equivalent if $\sigma^A(h) = \sigma^B(h)\, \forall h \in \mathcal{H}_{\omega\pi}$. 
We consider a basis set of policies for the surrogate approximate game $\Sigma_s$ of size $(|A|\cdot C)^D$, where $\forall h \in \mathcal{H}_{\omega\pi}, a \in A, \exists \sigma_s \in \Sigma_s \text{ s.t. } \sigma_s(h) = a$. For completeness each policy $\sigma_s \in \Sigma_s$ falls back to an arbitrary fixed default policy $\sigma_{\text{default}}$ for all $h \notin \mathcal{H}_{\omega\pi}$.

Given that the size of the true policy space is larger than the size of the surrogate policy space, we can construct a surjective mapping between the true policy space and the tree-equivalent surrogate policy space.
\begin{equation}
    \phi : \Sigma \rightarrow \Sigma_s
\end{equation}
\begin{equation}
    \phi(\sigma) = \sigma_s \in \Sigma_s \text{ s.t. } \sigma(h) = \sigma_s(h)\, \forall h \in \mathcal{H}_{\omega\pi}
\end{equation}

\begin{lemma} \label{lemma:tree-equiv-utility}
    Two tree-equivalent policies yield the same utility when evaluated with the same sparse-sampling-$\omega$-$\pi$ tree.
\end{lemma}
\begin{proof}
    

    By definition, two tree-equivalent policies ($\sigma_1, \sigma_2$) have $\sigma_1(h_t) = \sigma_2(h_t) \, \forall h_t \in \mathcal{H}_{\omega\pi}$.
    Again, by definition, a sparse-sampling-$\omega$-$\pi$ tree only evaluates histories $h_t \in \mathcal{H}_{\omega\pi}$.
    If the evaluated histories are identical, and the actions taken by policies in these histories are identical, then the policy values yielded by sparse-sampling-$\omega$-$\pi$ are also identical.
    
    

\end{proof}

\begin{lemma} \label{lemma:approx-dev-incentive}
    In a particle CDIT, a player has no incentive to unilaterally deviate to a strategy not in the surrogate policy space.
\end{lemma}
\begin{proof}
    (sketch) A strategy not explicitly considered by a sparse sampling-$\omega$-$\pi$ tree  ($\sigma$) has a tree-equivalent policy in the tree ($\sigma_s = \phi(\sigma)$) that, by definition, yields the same estimated utility. Because the estimated utilities are the same, there is no incentive to deviate from $\sigma_s$ to $\sigma$ in the approximate game. A more detailed proof is given in the appendix.
\end{proof}

\subsection{General Sum Particle CDIT Nash Equilibrium Approximation}

We are now ready to prove one of the main results of the paper: the bounded suboptimality of solving the approximate game yielded by a particle CDIT.

\begin{theorem} \label{theorem:arbitrary-policy-nashconv}
    The following bound holds with probability of at least $1 - 5\sum_{i \in \mathcal{N}}|\Sigma|(4C)^{D+1}\exp(-C \cdot \acute{k}_i^2)$:
    \begin{equation}
        \textsc{NashConv}^A(\pi) \le \sum_{i \in \mathcal{N}} \left[\delta_{i,s}^{\hat{A}^i}(\pi) + 2\epsilon_{\omega\pi}\right] ,
    \end{equation}
    where
    $\epsilon_{\omega\pi}, \acute{k}$ are defined in \cref{theorem:single-policy-error-bound}.
\end{theorem}
\begin{proof}
    From \cref{lemma:nashcov-bound}, we know that 
    \begin{equation}
        \textsc{NashConv}^A(\pi) \le \sum_i \left[\delta_{i,s}^{\hat{A}^i}(\pi) + 2||E^i||_\infty\right] \, ,
    \end{equation}
    where $\hat{A}^i$ is a game considering the full policy space.
    
    From \cref{lemma:approx-dev-incentive}, we find that $\delta_{i,s}^{\hat{A}^i}(\pi) = \delta_{i}^{\hat{A}^i}(\pi)$, indicating that we need only consider the smaller surrogate policy space $\Sigma_s$, leading to the following suboptimality bound
    \begin{equation}
        \textsc{NashConv}^A(\pi) \le \sum_{i \in \mathcal{N}} \left[\delta_{i,s}^{\hat{A}^i}(\pi) + 2||E^i||_\infty\right] \, .
    \end{equation}
    From \cref{theorem:single-policy-error-bound}, we know that utility approximation error for any policy $\sigma$ given the reward function for player $i$, and the rest of the constants follow directly from this theorem.

    While \cref{theorem:single-policy-error-bound} produces a probabilistic bound relevant to a single policy and a single player's reward function, this bound must now hold for all players and all possible joint policies. As such, a union bound is applied to the satisfaction probability of \cref{theorem:single-policy-error-bound} over all pure policies and all players, yielding a multiplicative factor of $|\Sigma|$ and a summation over all players $i \in \mathcal{N}$. Here $|\Sigma| = \sum_{d=0}^D (|\mathcal{A}|\cdot|\mathcal{O}|)^d$ is the number of possible joint closed-loop policies.
\end{proof}

Naturally, for \cref{theorem:arbitrary-policy-nashconv} to hold, the number of $D$-step policies must be finite, consequently requiring that both $\mathcal{A}$ and $\mathcal{O}$ also be finite sets.

\subsection{ESCFR Convergence in Zero Sum POSGs} \label{sec:proof-final-convergence}

For establishing more concrete bounds for zero-sum games, we demonstrate the full game convergence when using ESCFR over a particle CDIT.

\begin{theorem}
    For two-player zero-sum games, the following bound holds with probability of at least $1 - 2[p + 5|\Sigma|(4C)^{D+1}\exp(-C \cdot \acute{k}^2)]$:
    \begin{equation}
        \textsc{NashConv}^A(\pi) \le 
        4(\epsilon + \epsilon_{\omega\pi})
    \end{equation}
    where,
    \begin{equation}
        \epsilon = \left(1 + \frac{\sqrt{2}}{\sqrt{p}}\right)\frac{\max_i\left[\Delta_{u,i}\left[|\mathcal{A}^i|^3C\right]^{\frac{D+1}{2}}\right]}{2\sqrt{T}}\,,
    \end{equation}
    $\epsilon_{\omega\pi}, \acute{k}$ are defined in \cref{theorem:single-policy-error-bound}.

\end{theorem}
\begin{proof}
    For a given player $i$, average overall regret of ESCFR is bounded by
    \begin{equation}
        R_i^T \le \left(1 + \frac{\sqrt{2}}{\sqrt{p}}\right) \frac{\Delta_{u,i}M_i\sqrt{|A_i|}}{\sqrt{T}} \, ,
    \end{equation}
    with probability $1-p$ after $T$ ESCFR iterations~\cite{lanctot2009mccfr}.
    Here, $R_i^T$ denotes the bounded average overall regret, and $T$ denotes the number of CFR iterations performed. $\Delta_{u,i}$ denotes the range of terminal utilities to player $i$. For EFGs this utility is realized purely at the terminal histories of the game. However, for the POSGs that we consider, reward is accumulated, and the EFG-equivalent terminal utility is the discounted sum of expected rewards of a given history. $A_i$, in this case is the set of all player $i$ action history subsequences. For POSGs this is simply $\sum_{d=0}^D|\mathcal{A}^i|^d$. $M_i = \sum_{a_i \in A_i} \sqrt{|\mathcal{I}_i(a_i)|}$, where $\mathcal{I}_i(a_i)$ is the set of all information sets where player $i$'s action sequence up to that information set is $a_i$.

    The number of information sets where player $i$'s action sequence up to that information set is $a_i$ at depth $d$ is equivalent to the number of distinct possible observation sequences $C^d$.

    Thus,
    \begin{equation}
        M_i = \sum_{d=0}^D |\mathcal{A}^i|^d C^{d/2}\,.
    \end{equation}
    For the sake of simplicity, we crudely upper bound the finite geometric sum with an exponential term
    \begin{equation}
        \sum_{d=0}^D x^d = \frac{x^{D+1} - 1}{x - 1} \le x^{D+1} \quad  \forall x \ge 2 \, .
    \end{equation}
    With this simplification, we have
    \begin{equation}
    \begin{aligned}
        M_i\sqrt{|A_i|} &= \left[\sum_{d=0}^D |\mathcal{A}^i|^d C^{d/2}\right]\sqrt{\sum_{d=0}^D |\mathcal{A}^i|^d} \\
        &\le \left[ |\mathcal{A}^i|^{D+1} C^{(D+1)/2}\right]\sqrt{ |\mathcal{A}^i|^{D+1}} \\
        &\le \left[|\mathcal{A}^i|^3 C\right]^{\frac{D+1}{2}}\,,
    \end{aligned}
    \end{equation}
    with the added stipulation that $|\mathcal{A}^i| \ge 2$.

For the finite horizon problems we consider, $\Delta_{u,i}$ can be expressed as a finite geometric sum of rewards
\begin{equation}
\begin{aligned}
    \Delta_{R,i} &= \max_{s,a}R^i(s,a) - \min_{s,a}R^i(s,a)\\
    \Delta_{u,i} &= \frac{\Delta_{R,i}(1-\gamma^D)}{1-\gamma}
\end{aligned}
\end{equation}

In a zero-sum game, if $R_i^T \le \epsilon$, then average strategy $\bar{\sigma}^T$ is a $2\epsilon$ equilibrium defined by

\begin{equation}
    u_i(\sigma_i, \sigma_{-i}) \ge u_i(\sigma_i', \sigma_{-i}) - \epsilon,\, \forall i \in \mathcal{N}, \sigma_i' \in \Sigma_i\,.
\end{equation}

By definition of $\delta$ and $\epsilon$, we have $\forall i \in \mathcal{N}\, \delta^i \le \epsilon$ as well as $\epsilon \le \max_{i \in \mathcal{N}} \delta^i$.
    In a zero-sum game, if $R_i^T \le \epsilon$ for all players, then average strategy $\pi^T$ is a $2\epsilon$ equilibrium.
    Consequently, 
    \begin{equation}
        \epsilon \le \max_i R_i^T / 2 \le \left(1 + \frac{\sqrt{2}}{\sqrt{p}}\right)\frac{\max_i\left[\Delta_{u,i}\left[|\mathcal{A}^i|^3C\right]^{\frac{D+1}{2}}\right]}{2\sqrt{T}} \,,
    \end{equation}
    with probability $1 - 2p$ due to a union bound over regret bound satisfaction probability for both players.

    We combine this bound for $\delta^i$ specific to ESCFR with the suboptimality bound in \cref{theorem:arbitrary-policy-nashconv}. The satisfaction probability of this final bound is the result of a union bound over the satisfaction probabilities of ESCFR regret bounds and policy value error bounds.
\end{proof}

\section{Numerical experiments} \label{sec:experiments}

To demonstrate the effectiveness of our solver, we construct a game of tag in 2-dimenstional continuous state space for each agent (4 dimensions total) with discrete finite actions and observations.
A pursuer and an evader start in positions uniformly randomly sampled from $b_0 = \mathcal{U}[-0.25, 0.25] \times \mathcal{U}[-0.25, 0.25]$.
At each step, each agent observes the quadrant the other agent's relative position is in.
Both agents can move in 6 equally spaced directions. The pursuer only receives a reward of 1 once coming within a radius of 0.1 of the evader. As this is a zero-sum game, $r^{\text{pursuer}}(s,a) = -r^{\text{evader}}(s,a)$.

\begin{figure}
    \centering
    \includegraphics[width=\linewidth]{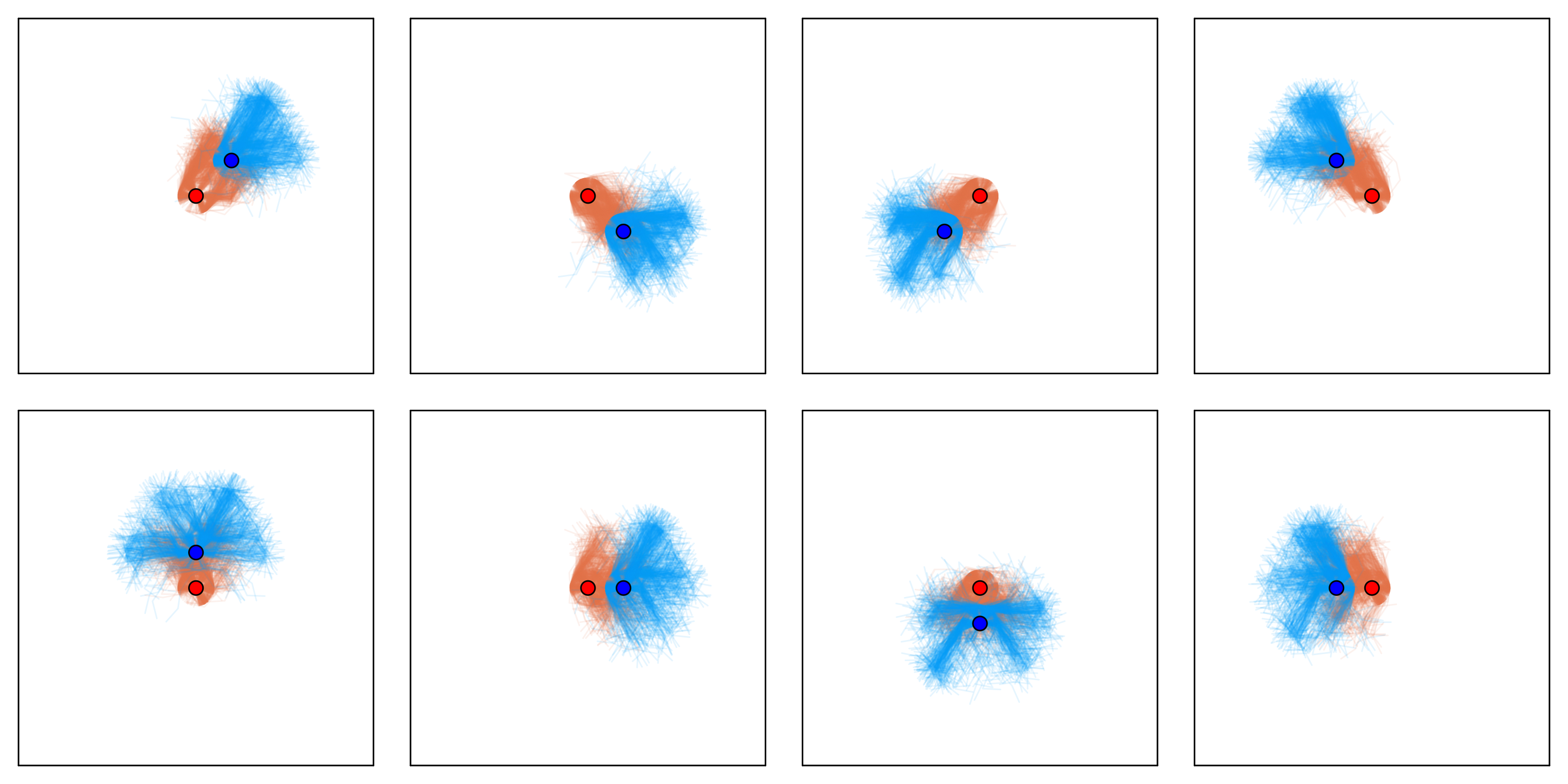}
    \caption{Continuous tag policies marginalized over observations}
    \label{fig:ctag-open-loop-plans}
\end{figure}

\begin{figure}
    \centering
    \includegraphics[width=\linewidth]{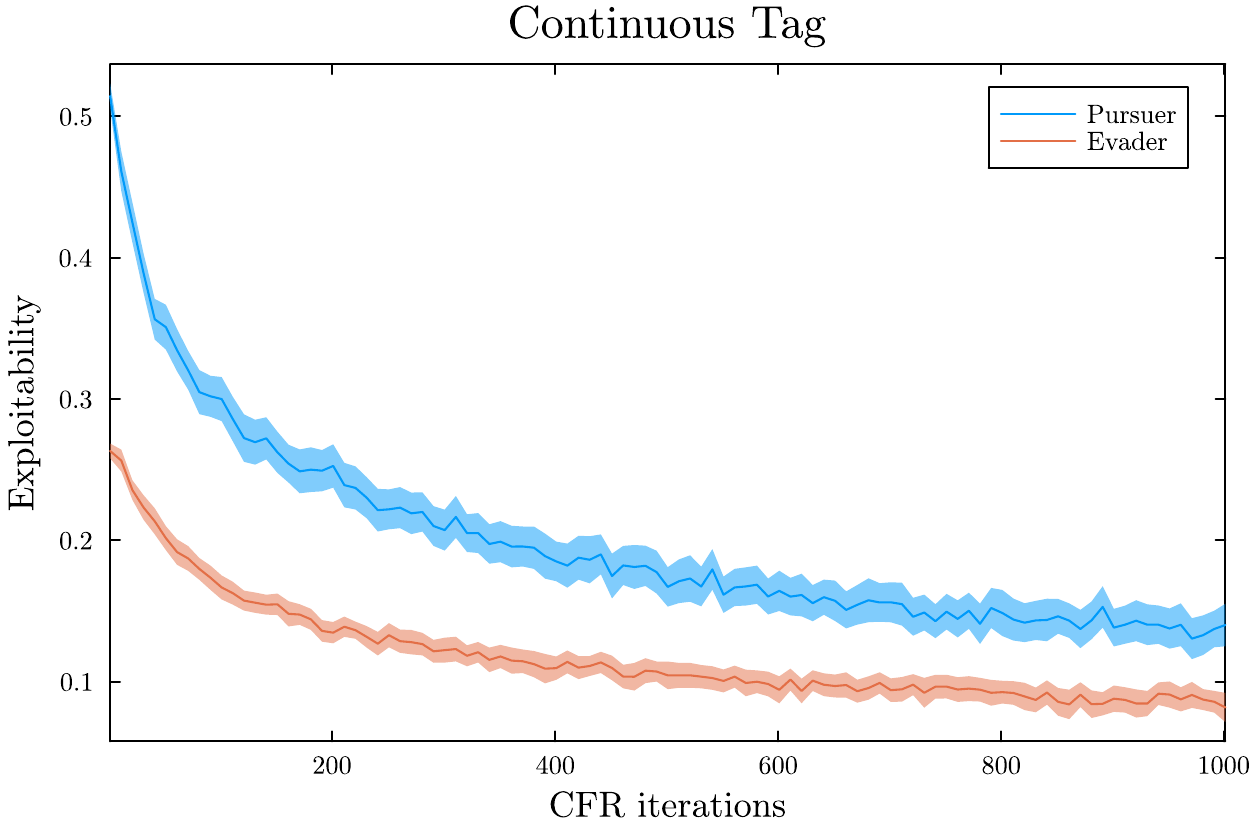}
    \caption{Continuous Tag exploitability; $3\sigma$ standard error bounds shaded}
    \label{fig:continuous_tag_exploitability}
\end{figure}



To elucidate the adversarial nature of the interaction uncertainty in these zero-sum games, we quantify suboptimality via exploitability, which we define to be how much utility an opponent is able to take away i.e. $e^i(\pi) = v^i(\pi) - \min_{\pi^{-i}}v^i(\pi^i, \pi^{-i})$. For Player $i$, this is equivalent to $\delta^{-i}$, and thus the sum of exploitabilities is equivalent to \textsc{NashConv} in two-player zero-sum games.


To solve the game of continuous tag, we construct and solve  a particle CDIT up to 5 time steps with beliefs consisting of 100 particles and 1000 total ESCFR iterations.
\Cref{fig:ctag-open-loop-plans} shows a policy found by ESCFR marginalized over all possible observations.
More opaque lines show more likely trajectories.
It is clear that this policy is highly stochastic, indicating that both agents are engaging in deception that would be impossible for a POMDP planner to choose.
To estimate exploitability of an ESCFR policy, POMCP~\cite{silver2010pomcp} is used as a best-responder.
Due to the stochasticity in tree construction and ESCFR solution, the game is solved 100 times to generate the mean and $3\sigma$ standard error exploitability curves displayed in \cref{fig:continuous_tag_exploitability}.

\section{Conclusion} \label{sec:conclusion}

This paper proposes a new approach for solving partially observable stochastic games by combining imperfect information game solution methodology with distribution approximations developed for POMDPs.
This combination advances the state-of-the-art in POSG solutions by extending solutions to continuous state spaces and large observation spaces, providing low-exploitability solutions with high probability.
While we provide a theoretical basis for sparse sampling methods in POSGs, this comes at the sacrifice of practical performance. ESCFR reduces the searched space to being purely exponential in a single agent's action space rather than exponential in the joint action and observation space, but this still becomes intractable for large action spaces and depths. Future work may consider model-free deep reinforcement learning methods to sample from the CDIT even more sparsely and generalize sampled results with deep learning. While the proposed methodology successfully handles continuous state spaces using sparse sampling, it currently operates without online replanning capabilities. However, extensions to the framework are feasible, drawing on approaches such as~\cite{sokota2023update,solinas2024history}, which provide avenues for integrating online strategy refinement.  




\bibliographystyle{ACM-Reference-Format} 
\bibliography{local_references, zach_references}

\newpage
\section{Appendix}
\subsection*{Proof of \cref{lemma:nashcov-bound}} \label{app:nashcov-bound-proof}
\begin{proof}
    \begin{equation}
    \begin{aligned}
    \delta_i^A(\pi) &= \max_{\pi_i^\prime}\left(\pi_i^{\prime T} A \pi_{-i}\right) - \pi_i^T A \pi_{-i} \\
    &= \max_{\pi_i^\prime}\left(\pi_i^{\prime T} (\hat{A} + E) \pi_{-i}\right) - \pi_i^T (\hat{A} + E) \pi_{-i} \\
    &\le \left\{\max_{\pi_i^\prime}\left(\pi_i^{\prime T} \hat{A} \pi_{-i}\right)
    - \pi_i^T \hat{A} \pi_{-i} \right\}
    + \left\{\max_{\pi_i^\prime}\left(\pi_i^{\prime T} E \pi_{-i}\right)
    - \pi_i^T E \pi_{-i} \right\}\\
    &= \delta_i^{\hat{A}}(\pi) + \delta_i^{E}(\pi) \\
    &\le \delta_i^{\hat{A}}(\pi) + 2||E||_\infty
    \end{aligned}
    \end{equation}

    Consequently,
    \begin{equation} 
        \textsc{NashConv}_A(\pi) \le \sum_i \left[\delta^i_{\hat{A}^i}(\pi) + 2||E^i||_\infty\right] .
    \end{equation}

\end{proof}

\subsection*{Proof of \cref{lemma:approx-dev-incentive}} 
\begin{proof}
Suppose we have a surrogate approximate game wherein player $i$ has desire $\delta^i$ to deviate. If we allow player $i$ the policy space afforded in the full approximate game, they will still only have desire $\delta^i$ to deviate.

\begin{equation}
    \delta^i_s = \max_{\sigma^i_s \in \Sigma_s} \left[\hat{U}^i(\sigma^i_s, \pi^{-i})\right] - \hat{U}(\pi^i, \pi^{-i})
\end{equation}

\begin{equation}
    \hat{U}^i(\sigma^i, \pi^{-i}) = \sum_{\sigma^{-i} \in \Sigma^{-i}} \hat{U}(\sigma^i, \sigma^{-i})\pi^{-i}(\sigma^{-i})
\end{equation}

\begin{equation}
    \delta^i = \max_{\sigma^i \in \Sigma} \left[\hat{U}^i(\sigma^i, \pi^{-i})\right] - \hat{U}(\pi^i, \pi^{-i})
\end{equation}

By consequence of \cref{lemma:tree-equiv-utility}, we get the following relation
\begin{equation}
\begin{aligned}
    \delta^i - \delta^i_s &= \max_{\sigma^i \in \Sigma} \left[\hat{U}^i(\sigma^i, \pi^{-i})\right] - \max_{\sigma^i_s \in \Sigma_s} \left[\hat{U}^i(\sigma^i_s, \pi^{-i})\right] \\
    &= \max_{\sigma^i \in \Sigma} \left[\hat{U}^i(\phi(\sigma^i), \pi^{-i})\right] - \max_{\sigma^i_s \in \Sigma_s} \left[\hat{U}^i(\sigma^i_s, \pi^{-i})\right] \\
    &= \max_{\sigma^i_s \in \Sigma_s} \left[\hat{U}^i(\sigma^i_s, \pi^{-i})\right] - \max_{\sigma^i_s \in \Sigma_s} \left[\hat{U}^i(\sigma^i_s, \pi^{-i})\right] \\
    &=  0 \, .\\
\end{aligned}
\end{equation}

Therefore, $\delta^i = \delta^i_s$

\end{proof}

\subsection*{Proof of \cref{theorem:single-policy-error-bound}} \label{app:single-policy-error-bound-proof}

\begin{theorem} \label{theorem:lev-2024} (Theorem 3 of \cite{lev2024simplifying})
Assume that the immediate state reward estimate is probabilistically bounded such that $P(|r_j^i - \tilde{r}_{j}^{i} | \ge \nu) \le \delta_r(\nu, N_r)$, for a number of reward samples $N_r$ and state sample $x_i^j$. Assume that $\delta_r(\nu, N_r) \rightarrow 0$ as $N_r \rightarrow \infty$. For all policies $\pi, t = 0, . . . , L$, and $a \in A$, the following bounds hold with probability of at least $1 - 5(4C)^{D+1}(\exp(-C \cdot \acute{k}^2) + \delta_r(\nu, N_r))$:
\begin{equation}
    \left|Q_{\mathbf{P}, t}^{\pi,\left[p_Z / q_Z\right]}\left(b_t, a\right)-Q_{\mathbf{M}_{\mathbf{P}}, t}^{\pi,\left[p_Z / q_Z\right]}\left(\bar{b}_t, a\right)\right| \leq \alpha_t+\beta_t = \epsilon,
\end{equation}
where
\begin{equation}
    \alpha_t=(1+\gamma) \lambda+\gamma \alpha_{t+1}, \, \alpha_L=\lambda \geq 0,
\end{equation}
\begin{equation}
    \beta_t=2 \nu+\gamma \beta_{t+1},\, \beta_L=2 \nu \geq 0
\end{equation}
\begin{equation}
    k_{\max }(\lambda, C)=\frac{\lambda}{4 V_{\max } d_{\infty}^{\max }}-\frac{1}{\sqrt{C}}>0,
\end{equation}
\begin{equation}
    \acute{k}=\min \left\{k_{\max }, \lambda / 4 \sqrt{2} V_{\max }\right\} .
\end{equation}
\end{theorem}
While \cite{lev2024simplifying} further generalize particle-belief MDP policy evaluation to simplified observation models, we assume that the observation model is known. This simplifies the R\'enyi divergence back to the definition provided in \cite{lim2023optimality}, where $\mathcal{P}^d$ is the target distribution and $\mathcal{Q}^d$ is the sampling distribution for particle importance sampling.

\begin{equation}
    d_\infty(\mathcal{P}^d||\mathcal{Q}^d) = \esssup_{x\sim\mathcal{Q}^d} w_{\mathcal{P}^d/\mathcal{Q}^d} (x) \le d_\infty^{\max} < +\infty
\end{equation}

In order to extend $V_{\max}$ from \cref{theorem:lev-2024} to the multiagent setting, we specify that $V_{\max, i}$ bounds value for player $i$ with a finite geometric sum such that
\begin{equation}
    V_{\max, i} = \frac{\max_{s,a}|R^i(s,a)| (1 - \gamma^D)}{1-\gamma} \, .
\end{equation}

Furthermore, we assume reward to be a deterministic function of a given state and action. As such, we have $\forall N_r > 0, \nu \ge 0,\, \delta_r(\nu, N_r) = 0$. Consequently, we can simple choose $\nu=0$ and $N_r=1$. 

By definition of $\beta_t$, we now have $\beta_t = 0\, \forall t$.

This reduces the concentration bound to
\begin{equation}
    \left|Q_{\mathbf{P}, t}^{\pi,\left[p_Z / q_Z\right]}\left(b_t, a\right)-Q_{\mathbf{M}_{\mathbf{P}}, t}^{\pi,\left[p_Z / q_Z\right]}\left(\bar{b}_t, a\right)\right| \leq \alpha_t = \epsilon,
\end{equation}
where
\begin{equation}
    \alpha_t=(1+\gamma) \lambda+\gamma \alpha_{t+1}, \, \alpha_D=\lambda \geq 0.
\end{equation}

By noting the sequence of $\alpha_0^D$ for different horizons $D$

\begin{equation}
\begin{aligned}
    \alpha^0_0 &= \lambda \\ 
    \alpha^1_0 &= \lambda(1 + 2\gamma) \\
    \alpha^2_0 &= \lambda(1 + 2\gamma + 2\gamma^2) \\
    &\vdots \\
    \alpha^D_0 &= \lambda(1 + 2 \sum_{t=1}^D\gamma^t), \\
\end{aligned}
\end{equation}

we can establish a closed form via finite geometric sum where
\begin{equation}
    \alpha_0 = \lambda\left[
    2\left(\frac{1-\gamma^{D+1}}{1-\gamma}\right) - 1
    \right],
\end{equation}
for a given problem horizon $D$.

We take the established guarantee at the root belief, and 
\begin{equation}
\begin{aligned}
    U^{i}(\sigma) &= Q_{\mathbf{P}, 0}^{\sigma,\left[p_Z / q_Z\right]}\left(b_0, \sigma(h_0)\right) \\
    \hat{U}^{i}(\sigma) &= Q_{\mathbf{M}_{\mathbf{P}}, 0}^{\sigma,\left[p_Z / q_Z\right]}\left(\bar{b}_0, \sigma(h_0)\right)\, ,
\end{aligned}
\end{equation}
for a given player $i$.

For all policies $\sigma, t = 0, . . . , D$, and $a \in A$, the following bounds hold with probability of at least $1 - 5(4C)^{D+1}\exp(-C \cdot \acute{k}_i^2)$:
\begin{equation}
    \left|U^{i}(\sigma)-\hat{U}^{i}(\sigma)\right| \leq \epsilon,
\end{equation}
where
\begin{equation}
    \epsilon = \lambda\left[
    2\left(\frac{1-\gamma^{D+1}}{1-\gamma}\right) - 1
    \right]
\end{equation}
\begin{equation}
    k_{\max, i}(\lambda, C)=\frac{\lambda}{4 V_{\max, i} d_{\infty}^{\max }}-\frac{1}{\sqrt{C}}>0,
\end{equation}
\begin{equation}
    \acute{k}_i=\min \left\{k_{\max, i}, \lambda / 4 \sqrt{2} V_{\max, i}\right\} .
\end{equation}

For zero-sum games, we have $R^i(s,a) = -R^{-i}(s,a)$. As such, $|R^i(s,a)| = |R^{-i}(s,a)|$, consequently allowing for the following equivalencies:
\begin{equation}
\begin{split}
    V_{\max} &= V_{\max, i} = V_{\max, -i} \\
    k_{\max} &= k_{\max, i} = k_{\max, -i} \\
    \acute{k} &= \acute{k}_{i} = \acute{k}_{-i} \\
\end{split}
\end{equation}

\end{document}